\documentclass[fleqn,usenatbib]{mnras}

\usepackage{newtxtext,newtxmath}

\usepackage[T1]{fontenc}

\DeclareRobustCommand{\VAN}[3]{#2}
\let\VANthebibliography\thebibliography
\def\thebibliography{\DeclareRobustCommand{\VAN}[3]{##3}\VANthebibliography}

\usepackage{graphicx}	
\usepackage{amsmath}	
\usepackage{amssymb}	

\title[Time lags of the type-B Quasi-Periodic Oscillation in MAXI J1348-630]
{Time lags of the type-B QPO in MAXI J1348-630}

\author[T.M. Belloni]{
Tomaso M. Belloni$^{1}$\thanks{E-mail: tomaso.belloni@inaf.it},
Liang Zhang $^{2}$,
Nikolaos D. Kylafis$^{3,4}$,
Pablo Reig$^{4,3}$ \&
\newauthor
Diego Altamirano$^{2}$
\\
$^{1}$INAF-Osservatorio Astronomico di Brera, via E. Bianchi 46, I-23807, Merate, Italy\\
$^{2}$Physics and Astronomy, University of Southampton, Southampton, Hampshire SO17 1BJ, United Kingdom\\
$^{3}$University of Crete, Physics Department \& Institute of Theoretical \& Computational Physics, 70013 Heraklion, Crete, Greece \\
$^{4}$Institute of Astrophysics, Foundation for Research and Technology-Hellas, 71110 Heraklion, Crete, Greece\\
}

\date{Accepted 2020 June 22. Received 2020 June 22; in original form 2020 May 11}

\pubyear{2020}

\begin{document}
\label{firstpage}
\pagerange{\pageref{firstpage}--\pageref{lastpage}}
\maketitle

\begin{abstract}
The fast variability observed in the X-ray emission from black-hole binaries has a very complex phenomenology, but offers the possibility to investigate directly the properties of the inner accretion flow. In particular, type-B oscillations in the 2-8 Hz range, observed in the Soft-Intermediate state, have been associated to the emission from a relativistic jet. We present the results of the timing and spectral analysis of a set of observations of the bright transient MAXI J1348-630 made with the NICER telescope. The observations are in the brightest part of the outburst and all feature a strong type-B QPO at $\sim$4.5 Hz. We compute the energy dependence of the fractional rms and the phase lags at the QPO frequency, obtaining high signal-to-noise data and sampling for the first time at energies below 2 keV. The fractional rms decreases from more than 10\% at 9 keV to 0.6\% at 1.5 keV, and is constant below that energy. Taking the 2-3 keV band as reference, photons at all energies show a hard lag, increasing with the distance from the reference band. The behaviour below 2 keV has never been observed before, due to the higher energy bandpass of previous timing instruments. The energy spectrum can be fitted with a standard model for this state, consisting of a thin disc component and a harder power law, plus an emission line between 6 and 7 keV. We discuss the results, concentrating on the phase lags, and show that they can be interpreted within a Comptonization model.
\end{abstract}

\begin{keywords}
accretion, accretion discs -- stars: black holes -- stars: jets -- X-rays: binaries -- X-rays: individual: GRS 1915+105
\end{keywords}



\section{Introduction}

Black-hole binaries show a complex and variable phenomenology in their X-ray light curves on sub-second  timescales \citep{BelloniStella2014,BelloniMotta2016,IngramMotta2020}. In particular, low-frequency Quasi Periodic Oscillations (QPO) are observed in the 0.1-30 Hz range. They have been divided into three types \citep[e.g.][and references therein]{Wijnands1999,Remillard2002,Casella2005,HomanBelloni2005,IngramMotta2020}.
Type-C QPOs always appear together with band-limited noise in the Power Density Spectra (PDS) and have been associated to Lense-Thirring precession in the inner region of the accretion flow \citep[see, e.g.][]{StellaVietri1998,Ingram2009,IngramMotta2020}. Type-A QPOs are weak, broad and elusive and for these reasons have not been extensively studied.
Type-B QPOs appear during the so-called Soft-Intermediate State (SIMS), a short-lived state in the evolution of black-hole transients where the energy spectrum is soft, dominated by an accretion disc component, but with a strong hard component \citep{Remillard2002,StevensUttley2016,BelloniMotta2016}. They have a typical centroid frequency around 4-6 Hz, although lower values have been observed, a rather high amplitude (fractional rms of a few \%) and are narrow, with a quality factor around 6 or higher \citep{Takizawa1997,Wijnands1999b,Remillard2002,Nespoli2003,Casella2005,IngramMotta2020}. Fast transitions between type-B QPOs and other types are often seen (see e.g. \citealt{Casella2004}). The fact that these transitions take place in rough correspondence of discrete jet ejections suggests that they are associated to the jet component \citep{FenderBelloniGallo2004}, but the precise association is not clear, as the time correspondence does not appear to be exact \citep{FenderHomanBelloni2009,Russell2020, Homan2020}. 
Explanations of phase lags in QPOs have been offered by \citet{Shaposhnikov2012} and \citet{MisraMandal2013}.
\cite{KylafisReig2019} offered a quantitative explanation of the type-B QPOs in GX 339-4, as coming from a precessing jet.  Their model assumes that the Comptonization takes place in the jet.  Because of the semi-relativistic bulk motion of the electrons in the jet, the upscattered photons exhibit a harder spectrum (lower power-law index $\Gamma$) along the jet than perpendicular to it.  In other words, $\Gamma$ increases with viewing angle with respect to the jet axis.  As a result, from a precessing jet, the observer sees a periodic variation of $\Gamma$. The period of variation is the period of the QPO, because the jet is thought to be fed from the precessing hot inner flow. Jet precession driven by inner-flow precession is also seen in GRMHD simulations \citep{Liska2018}.

MAXI J1348--630 is a bright X-ray transient discovered by the MAXI instrument on board the International Space Station (ISS) on 26 January 2019 \citep{yatabe2019,Tominaga2020}. It was detected by the Fermi satellite and by the Swift Burst Alert Telescope as a possible gamma-ray burst \citep{GCN1,GCN2}. Observations with the X-Ray Telescope (XRT) on board Swift yielded a precise position that confirmed the optical counterpart reported by \citep{Denisenko}. The first NICER observations suggested that the source is a probable black-hole binary, based on the properties of energy spectra and fast time variability \citep{Sanna2019}. 
\citet{Zhang2020} performed a detailed analysis of the X-ray spectral and variability study of all available NICER observations and concluded that MAXI J1348--630 hosts a black hole.
A spectral softening was detected by MAXI around 2019 February 3 \citep{Nakahira2019}, confirmed by INTEGRAL \citep{Cangemi2019}, indicating a state transition. MeerKAT radio observations showed a strong radio flare after the reported state transition \citep{Carotenuto2019}. Here we report about NICER observations made after the transitions, concentrating on those containing a type-B QPO. 

\section{Observations and data analysis}

MAXI J1348-630 was observed with the Neutron Star Interior Composition Explorer (NICER) on board the ISS since 26 January 2019, the day of its discovery. 
The X-Ray Timing Instrument (XTI) of NICER consists of 56 X-ray optics with silicon detectors and provides high time resolution data in the 0.2-12 keV energy range \citep{Gendreau2012}. Presently, 52 intruments were active and we removed detectors 13 and 14 as they occasionally show high electronic noise.  We made use of the public data in NASA's HEASARC archive. A total of 199 observations were made from 2019 January 26 through 2020 February 7, with exposures ranging from less than 100 s to 17.7ks.
We processed the original data using the NICERDAS pipeline, applying single pipeline elements in a sequence not to exclude the offset observations.

\begin{figure}
	\includegraphics[width=\columnwidth]{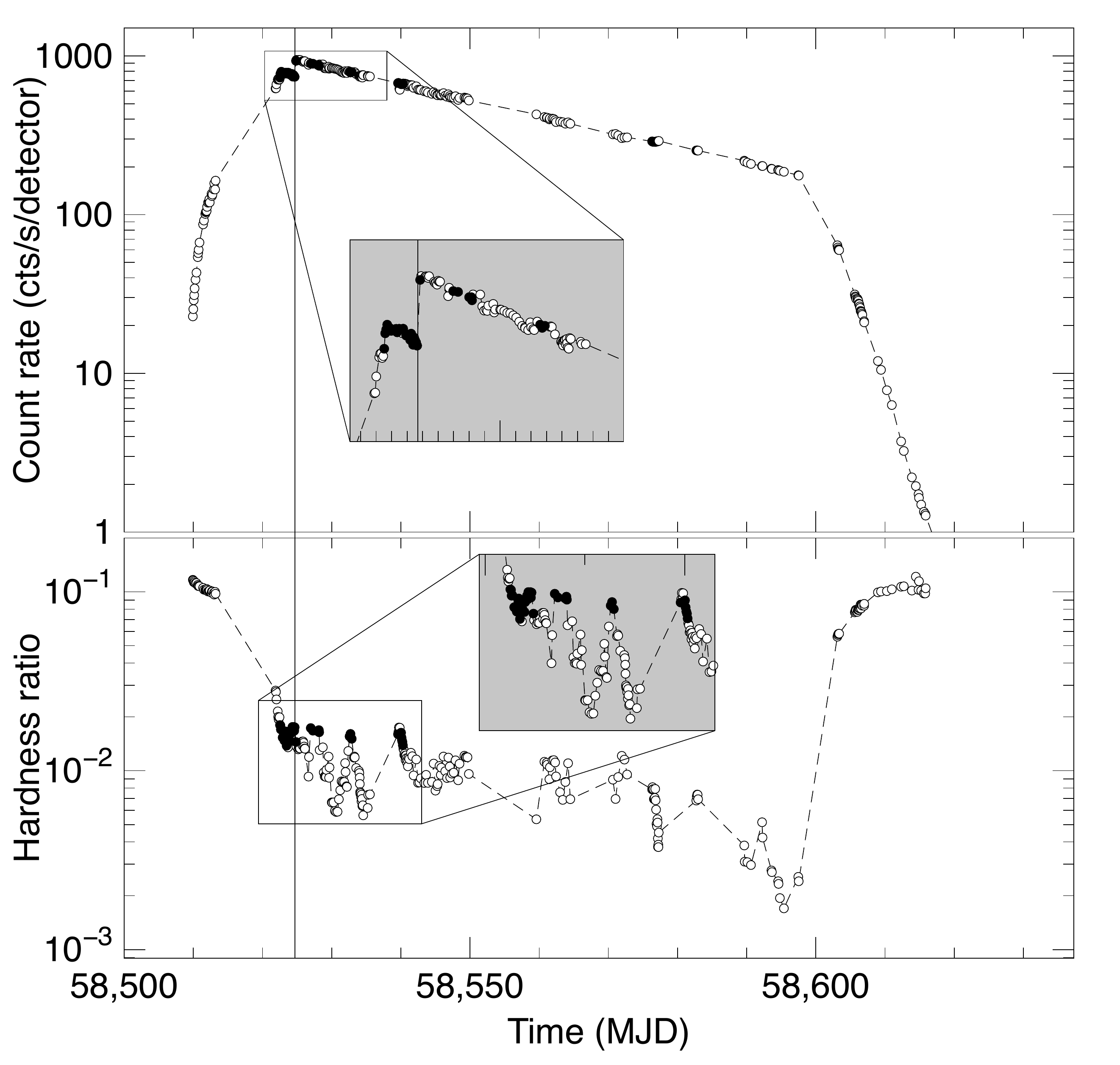}
    \caption{Top panel: net 0.5-12 keV light curve of the first part of the outburst of NICER J1348-630, one point per ISS orbit. The black points correspond to the data presented here, where a type-B QPO is present. The inset shows a magnification of the region containing the black points. The vertical line corresponds to the time of change in NICER pointing (see text).
    Bottom panel: corresponding time evolution of the (6-12 keV)/(2.0-3.5 keV) hardness ratio.
    }
    \label{fig:lightcurve}
\end{figure}

\begin{figure}
	\includegraphics[width=\columnwidth]{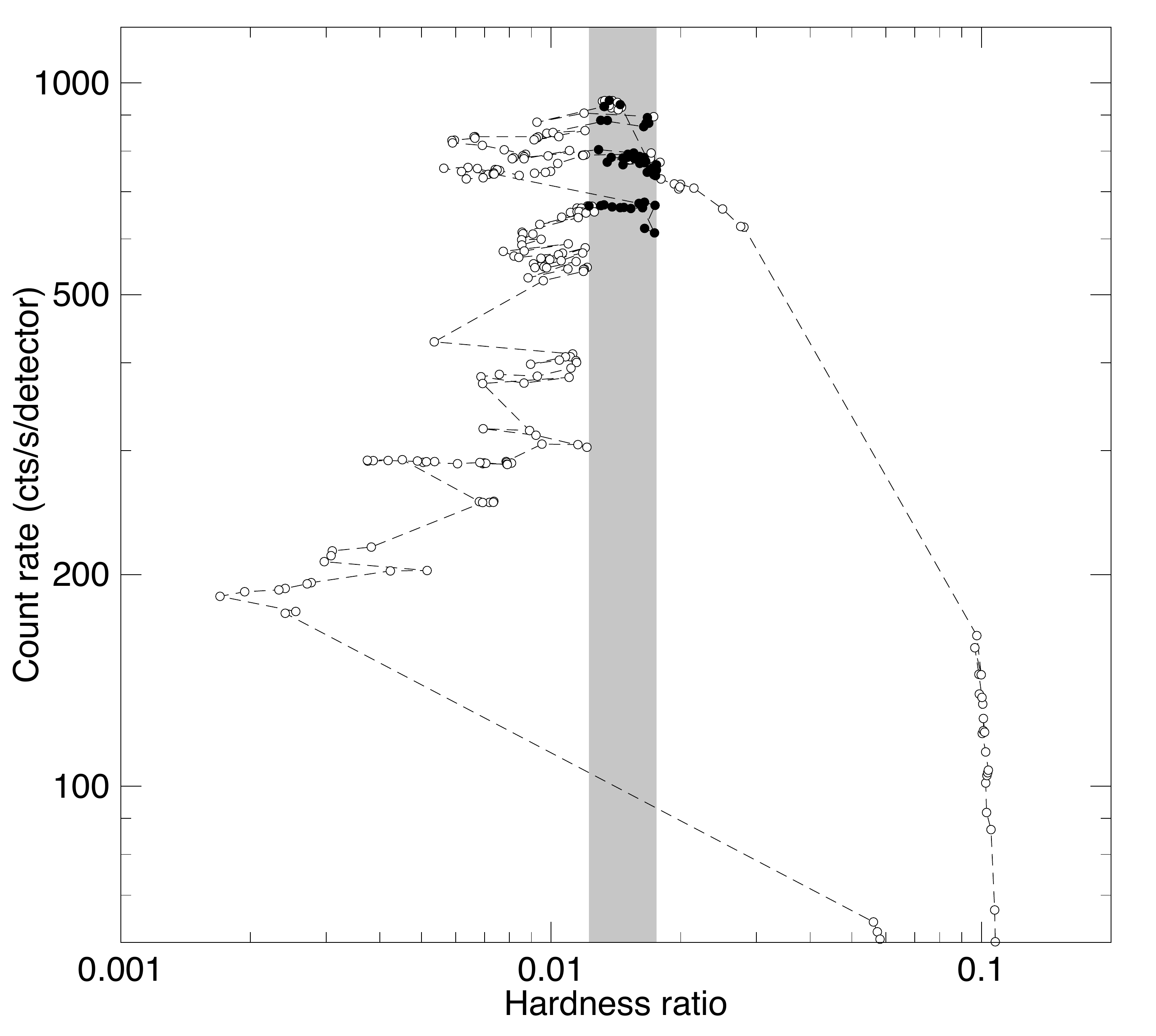}
    \caption{Hardness-Intensity diagram for all points shown in Fig. \ref{fig:lightcurve} with count rate larger than 60 cts/s/det. The black points correspond to the data presented here, where a type-B QPO is present. The gray band shows the range in hardness spanned by the black points.
    }
    \label{fig:hid}
\end{figure}

We extracted a net light curve in the 0.5-12 keV band to follow the outburst of the source using XSELECT (see top panel of Figure \ref{fig:lightcurve}, where each point corresponds to data from an ISS orbit). The background was calculated using the "nibackgen3C50" tool\footnote{\url{ https://heasarc.gsfc.nasa.gov/docs/nicer/tools/nicer_bkg_est_tools.html}}. The light curves were normalised by the number of active detectors. Early observations had a pointed offset which affects the count rate and spectral shape observed \citep[see][for more details]{Zhang2020}. Other observations have a reduced number of active instruments in order to reduce instrumental effects due to the  very high count rate observed. 
In Fig. \ref{fig:lightcurve} we show count rate  per detector, while the change in pointing direction can be seen as a sharp increase in count rate marked by a vertical dot-dash line. 
The mispointing had the effects of reducing the count rate and of introducing a quasi-periodic peak around 0.22 Hz, well below the frequency region of interest for this work and two additional spurious peaks at $\sim$1.5 Hz and $\sim$2 Hz in the Power Density Spectra (PDS), which we ignored in the analysis. They are due to the pointing offset that affects some observations. The full outburst continued after the range shown here, but we concentrate on the first part.
We also extracted net light curves in two energy bands, 2-3.5 keV and 6-12 keV, from which we obtained a hardness ratio, shown in the bottom panel of Fig. \ref{fig:lightcurve}. The Hardness-Intensity Diagram (HID) for the data above 60 cts/s/detector is shown in Fig. \ref{fig:hid}.
From Figs. \ref{fig:lightcurve} and \ref{fig:hid} it appears clear that a state transition took place between the first observations before MJD 58520 and those after that date, although unfortunately there is a one-week gap where most of the transition took place. From the evolution in the HID, it appears that the observations before MJD 58520 were made during the Low-Hard State (LHS) and at least the first white points in the figure after the gap were probably in the Hard-Intermediate State (HIMS), but only an inspection of the fast-timing properties can clarify that,

\subsection{Timing analysis}
For each observation, we extracted PDS from intervals of duration 13.1072 seconds over the energy range 0.2-12 keV. A number of observations show a clear type-B QPO in the PDS. Upon inspection, not all NICER orbit intervals in them contains a type-B, indicating fast state transitions \citep[see][]{Zhang2020}. We eliminated all PDS belonging to those orbits, and also checked for background flares by examining the 12-15 keV light curve at 1s binning. We eliminated all PDS including a bin with 12-15 keV count rate higher than 2 cts/s. Finally, we averaged the PDS belonging to each observation. The PDS were normalised after \citet{Leahy1983} and extended to a Nyquist frequency of 1250 Hz. The NICER orbits containing observations with a type-B QPO are marked with black points in Figs. \ref{fig:lightcurve} and \ref{fig:hid} and the observation log is reported in Tab. \ref{tab:observation_log}. The type-B QPOs appear at the local peak of the light curve, which is similar to what was found in GX 339-4 and XTE J1859+226 \citep{Casella2004,Motta2011}.
It is evident that all the eight observations  listed in Tab. \ref{tab:observation_log} correspond to a narrow range in hardness, which we therefore identify as that corresponding to the SIMS. A few data points in Fig. \ref{fig:hid} do not contain a type-B QPO, but are in the SIMS hardness range.

\begin{table}
	\centering
	\caption{Log of the NICER observations analyzed in this work. The third column reports the number of 13.1072s data intervals included in the analysis.}
	\label{tab:observation_log}
	\begin{tabular}{lccc} 
		\hline
		ObsId & Start date/time & N. Intervals& QPO Freq. (Hz)\\
		\hline
		1200530107 & 2019 Feb 08 00:08:53 &  86 &4.514$\pm$0.005\\
		1200530108 & 2019 Feb 09 03:57:20 & 283 &4.536$\pm$0.009\\
		1200530109 & 2019 Feb 10 00:02:40 & 551 &4.631$\pm$0.007\\
		1200530111 & 2019 Feb 11 23:57:21 &  27 &4.618$\pm$0.048\\
		1200530112 & 2019 Feb 13 02:15:00 & 172 &4.491$\pm$0.010\\
		1200530113 & 2019 Feb 13 23:54:20 &  75 &4.450$\pm$0.013\\
		1200530117 & 2019 Feb 18 00:08:00 & 200 &4.324$\pm$0.010\\
		1200530124 & 2019 Feb 25 01:37:44 & 261 &4.346$\pm$0.010\\
		1200530125 & 2019 Feb 26 00:41:12 & 276 &4.095$\pm$0.016\\
		\hline
	\end{tabular}
\end{table}

In order to obtain the QPO centroid frequency, we fitted the 3-20 Hz frequency range with a model consisting of a power law component for the noise, two Lorentzian peaks for the QPO and its overtone and a constant for the Poissonian noise. We limited the fit to above 3 Hz in order to avoid spurious peaks and complications in the noise shape and extended it until 500 Hz to have a good sampling of the Poissonian noise. The best fit centroid frequencies for the fundamental are shown in Tab. \ref{tab:observation_log}. They are all very similar, with the exception of observation 1200530125, where it is lower. A posteriori we found no significant difference removing it from further analysis, therefore we kept it in the sample.

\begin{figure}
	\includegraphics[width=\columnwidth]{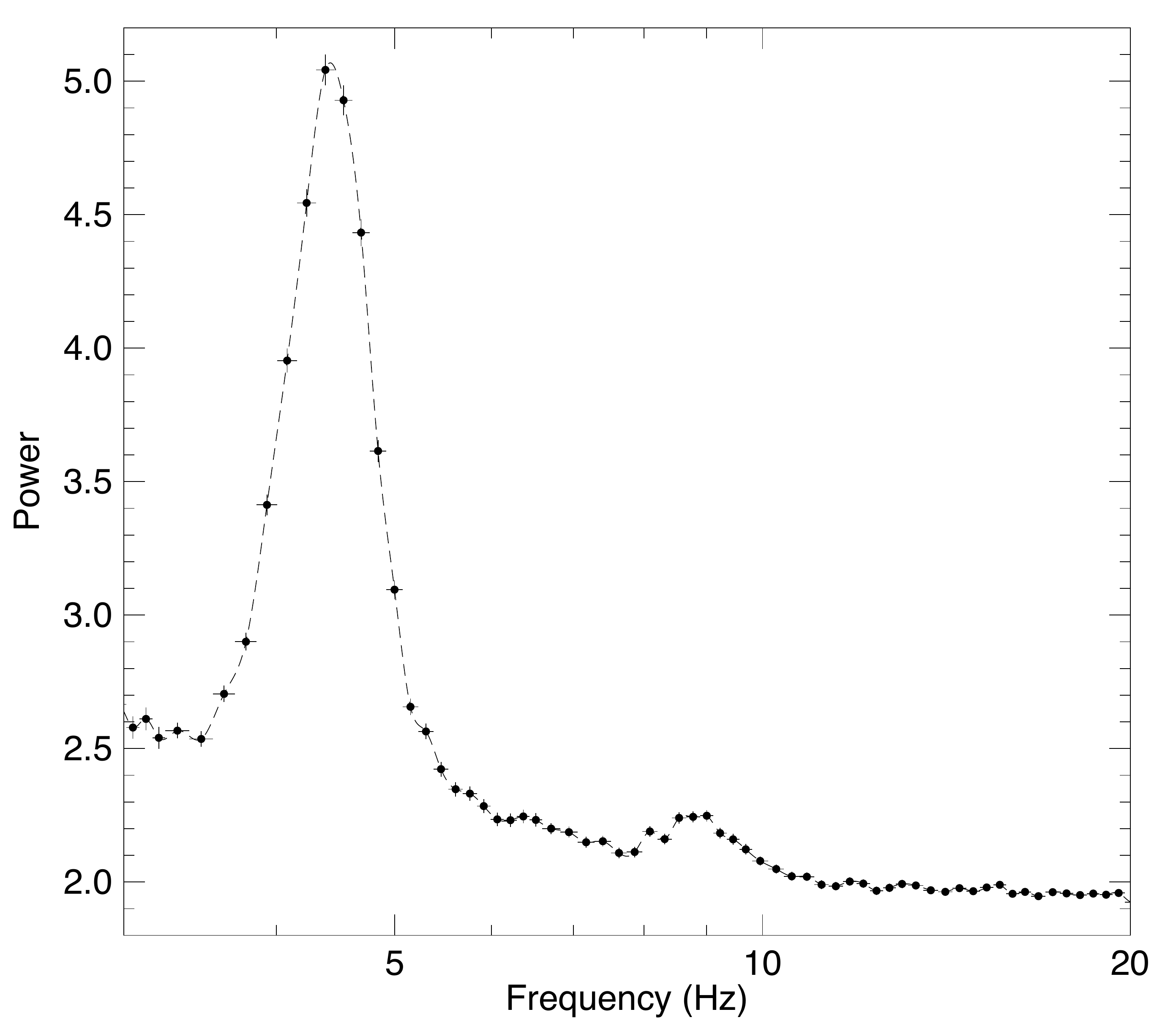}
    \caption{Leahy-normalised PDS obtained averaging data in the eight observations listed in Tab. \ref{tab:observation_log}.
    }
    \label{fig:average_pds}
\end{figure}

Since the QPO peaks appear to be very similar in frequency, we extracted PDS adding all observations together and obtained a final PDS summing 1931 individual PDS, for a total of 25310s of exposure. The average PDS is shown in Fig. \ref{fig:average_pds}.

\begin{figure}
	\includegraphics[width=\columnwidth]{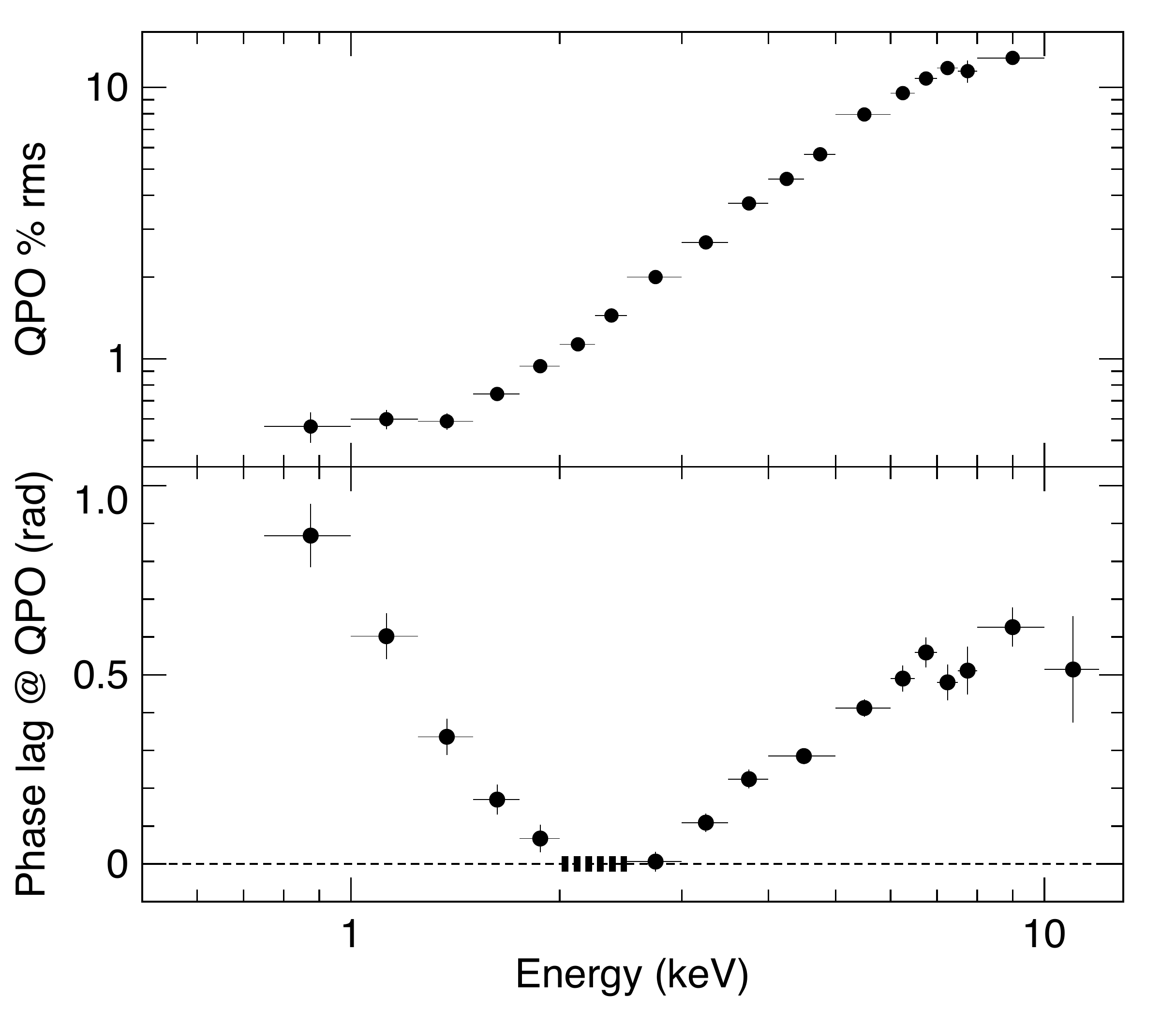}
    \caption{Top panel: QPO fractional rms as a function of energy. Bottom panel: corresponding phase lags at the QPO (see text). The reference band is shown by the dashed band.
    }
    \label{fig:rms_lag}
\end{figure}

A fit to the average PDS yields a centroid frequency of 4.475$\pm$0.006 Hz and a FWHM of 0.310$\pm$0.010 Hz. In order to examine the energy dependence of the QPO, we extracted PDS following the recipe above for different energy ranges and fitted them to obtain the fractional rms as a function of energy. Given the good statistics due to the intensity of the signal, we could obtain good PDS for 19 energy intervals, with a higher resolution in the 5-7 keV range. The QPO fractional rms as a function of energy is shown in the top panel of Fig. \ref{fig:rms_lag}. The background contribution to the count rate was estimated from the same background files used for spectral analysis. Background was of the order of 1 ct/s in all bands and was essentially negligible besides in the highest-energy band, which resulted in a meaningless upper limit for that point.

For 18 of the 19 energy intervals (the two intervals between 2.0 and 2.5 keV were merged to obtain a highr statistics) we extracted FFTs with the same procedure used for PDS and produced 17 cross spectra using one of the FFTs as a reference band. The reference band was chosen to be 2.0-2.5 keV. We computed the phase lags at the QPO frequency by integrating the cross spectra over a frequency band centred on the centroid frequency and one FWHM wide. The correction needed to take into account cross-channel talk was included, but found to be almost negligible. 
The lag vs. energy spectrum is shown in the bottom panel of Fig. \ref{fig:rms_lag}. With our choice of reference band, made a posteriori, phase lags are positive both below and above 2-3 keV. Notice that for a 4.45 Hz QPO a phase lag of 1 radian corresponds to a time lag of 36 ms.

\subsection{Spectral analysis}

The first three observations in our sample corresponded to the part of the outburst when NICER was not pointed directly to the source, therefore we did not consider them for spectral analysis. We extracted background-subtracted energy spectra using "nibackgen3C50" tool obtaining one spectrum per observation.
We fitted the 1-10 keV spectra with a model consisting of a disc blackbody and a Comptonised component, plus an emission line at $\sim$6.5 keV, all modified by interstellar absorption. This was achieved with the XSPEC model {\tt TBfeo*(diskbb*simpl+gaussian)}. The best fit parameters were similar for all observations, although in some case formally not compatible (see Tab. \ref{tab:spectra}). In all spectra there is an excess around 1.3 keV. A similar excess was found in the NICER spectra of Ser X-1 and was found not to be instrumental, but intrinsic to the source spectrum \citep{Ludlam2018}. We did not attempt to fit it as a detailed emission line modeling is beyond the scope of this work.

The 1-10 keV flux is dominated by the disc component, which has an inner disc temperature of $\sim$0.6 keV and a normalization that for a distance of 10 kpc and an inclination of 60 degrees translates into an inner disc radius of 250-350 km. The power-law component is steep, with a photon index $\Gamma$ around 3.5. The emission line is not always very significant, but is required by the fits. We remark that this is a simple spectral fitting done to confirm the presence of the SIMS: the derived inner radius is to be considered a very rough estimate. 

\begin{table*}
	\centering
	\caption{Best fit spectral parameters to the six observations without pointing offset. The parameters are: absorption column (in $10^{22}$ cm$^{-2}$), photon index and scattering fraction for the Comptonised component, inner disc temperature (in keV) and normalization for the disc blackbody, centroid energy (in keV), sigma (in keV) and normalization for the Gaussian emission line, best-fit chi square, 1-10 keV fluxes for the three components (in erg cm$^{-2}$s$^{-1}$)}
	\label{tab:spectra}
	\begin{tabular}{lcccccc} 
		\hline
		Parameter   & 1200530111        & 1200530112        & 1200530113        & 1200530117        & 1200530124        & 1200530125        \\
		\hline
		N$_H$       & 0.631 $\pm$ 0.003 & 0.628 $\pm$ 0.003 & 0.627 $\pm$ 0.002 & 0.630 $\pm$ 0.003 & 0.623 $\pm$ 0.003 & 0.626 $\pm$ 0.003 \\
		\hline
		$\Gamma$    & 3.58  $\pm$ 0.02  & 3.52  $\pm$ 0.03  & 3.54  $\pm$ 0.03  & 3.52  $\pm$ 0.04  & 3.35  $\pm$ 0.02  & 3.47  $\pm$ 0.02  \\
		Frac.       & 0.47  $\pm$ 0.03  & 0.40  $\pm$ 0.02  & 0.42  $\pm$ 0.02  & 0.36 $\pm$ 0.02  & 0.35  $\pm$ 0.01  & 0.32  $\pm$ 0.01  \\
		\hline
		kT$_{DBB}$  & 0.621 $\pm$ 0.009 & 0.642 $\pm$ 0.005 & 0.634 $\pm$ 0.007 & 0.631 $\pm$ 0.005 & 0.608 $\pm$ 0.003 & 0.614 $\pm$ 0.003 \\
		N$_{DBB} $  & (5.37$\pm$0.30)$\times 10^4$& (5.06$\pm$0.15)$\times 10^4$& (5.18$\pm$0.20)$\times 10^4$& (3.96$\pm$0.11)$\times 10^4$&  (4.94$\pm$0.11)$\times 10^4$&  (4.53$\pm$0.09)$\times 10^4$ \\
		\hline
		E$_{line}$  & 6.57  $\pm$ 0.08  & 6.65  $\pm$ 0.05  & 6.54  $\pm$ 0.07  & 6.59  $\pm$ 0.09  & 6.48  $\pm$ 0.04  & 6.56  $\pm$ 0.06  \\
	$\sigma_{line}$ & 0.11  $\pm$ 0.06  & 0.22  $\pm$ 0.06  & 0.16  $\pm$ 0.09  & 0.26  $\pm$ 0.17  & 0.45  $\pm$ 0.05  & 0.42  $\pm$ 0.05  \\
	    \hline
	$\chi^2 (dof)$  & 261 (216)         & 242 (216)         & 269 (216)         & 225 (216)         & 269 (216)         & 274 (216)         \\
	    \hline
	    F$_{DBB}  $   & $9.58\times 10^{-8}$ & $1.06\times 10^{-7}$ & $1.02\times 10^{-7}$ & $7.61\times 10^{-8}$ & $7.96\times 10^{-8}$ & $7.70\times 10^{-8}$ \\
	    F$_{Simpl}$   & $4.62\times 10^{-8}$ & $4.30\times 10^{-8}$ & $4.40\times 10^{-8}$ & $2.79\times 10^{-8}$ & $3.24\times 10^{-8}$ & $2.60\times 10^{-8}$ \\
	    F$_{Line}$   & $5.68\times 10^{-11}$ & $9,72\times 10^{-11}$ & $6.88\times 10^{-11}$ & $6.28\times 10^{-11}$ & $1.55\times 10^{-10}$ & $1.27\times 10^{-10}$ \\
		\hline
	\end{tabular}
\end{table*}

\section{Discussion and conclusions}

We have extracted the rms spectrum and the phase/time lag spectrum for the type-B QPO from a set of observations of MAXI J1348-630. Above 2 keV, the behavior of both spectra is consistent with what was observed earlier from type-B QPOs in different systems: the fractional rms increases with energy from below 1\% to more than 10\%, while the photons above 2 keV lag those at 2 keV, with a lag increasing with the considered energy band \citep[see e.g.][]{StevensUttley2016}.
Our results are consistent with previous observations of other systems, but given the effective area of NICER and the long exposure time we have very small uncertainties and can extend the energy coverage below 2 keV. We can see that the QPO fractional rms levels at $\sim$0.6\% below $\sim$1.5 keV. Surprisingly, the phase lags versus the 2-2.5 keV photons increase towards lower energies, reaching almost 0.9 rad (corresponding to a time lag of 32 ms) below 1 keV. The energy spectrum for different observations in our sample are rather typical of the SIMS, featuring a dominant accretion disc component and a steep hard component.
Notice that there is a significant feature in the phase-lag spectrum corresponding to the energy of the iron emission line. A full analysis and discussion of the line properties is not within the scope of this paper, but we note that it will be worth to analyse its possible variation with the QPO phase, as detected for a type-C QPO in H 1743-322 \citep{Ingram2016}. Phase-resolved spectroscopy was  performed on the type-B QPO of GX 339-4, but the statistics was not sufficient to resolve variations of the iron line \citep{StevensUttley2016}.

Time lags in the broad-band noise of black-hole binaries are thought to come from propagating mass-accretion-rate fluctuations \citep{Leahy1983,Kotov2001,Arevalo2006,Rapisarda2017}. In this picture, the fluctuations propagate into progressively hotter parts of the corona and the harder X-ray photons are produced {\it later} than the softer ones. Thus, there is a {\it positive} lag of the harder photons (say 10 keV) with respect to softer ones (say 2-3 keV), which are taken as a reference.  Naturally then,  softer energies below the reference band (say 1 keV) should exhibit {\it negative} lags.  This is the opposite of what we have observed in the QPO, indicating that type-B QPOs are not driven by mass-accretion-rate fluctuations (Fig. \ref{fig:rms_lag}).

Another mechanism that produces hard lags is Comptonization of soft photons in a hot corona. This is the preferred mechanism that is usually invoked to explain the power-law hard X-ray spectra in black-hole binaries \citep{Gierlinski1997,Gierlinski1999,PoutanenVurm2009,Done2007}. On average, soft photons gain energy from the hot electrons and, as a result of their random walk in the hot medium, more energetic photons come out {\it later} than the reference ones, which either have not scattered at all or have had their energy only slightly increased. However, in a Comptonization process as the one described above, not all input photons gain energy. Some of them lose energy with consecutive scatterings and because of their random walk in the hot medium come out later than the input photons.  In other words, if we consider the reference photons as having zero lag, then both higher and lower energy photons will exhibit positive lags, because they must random walk in the medium in order to have their energy significantly changed from its input value.  

Type-B QPOs are suspected to come from a precessing jet. Oscillations of the power-law index along the QPO oscillations have been mesasured in GX 339-4 \citep{StevensUttley2016}.  Recently, it was shown quantitatively that, if Comptonization takes place in the jet, i.e. if the jet serves the role of the hot corona, then its precession can explain quantitatively the variation of the power-law index $\Gamma$ of GX 339-4 with QPO phase  \citep{KylafisReig2019,KylafisReigPapadakis2020}.

It is beyond the scope of the present paper to show that Comptonization in a jet can reproduce quantitatively the observed time lags at both high and low energies.  What we can do is a simple, illustrative calculation using a spherical corona.  With a Monte Carlo code, we have computed the time lag of the Comptonised photons as a function of their energy.  The radius of the corona is $R = 10^9$ cm, its optical depth to Thomson scattering is $\tau = 0.3$, and the electron temperature in the corona is $kT_e = 30$ keV. A point source, with a flat spectrum between 2 and 3 keV and no emission at other energies, has been placed at the centre of the corona. 
Care has been taken to compute the last part of the flight of the photons accurately.  Thus, after the last scattering, the distance between the point of last scattering and the tangent plane to the sphere, that is perpendicular to the direction of the photon, is added to the distance traveled by the photon in its random walk.  This secures that all the photons that escape in the same direction {\it come in step} at the tangent plane.  After that, all of them travel the same distance. The time of flight $R/c$, where $c$ is the speed of light, of the photons that escaped unscattered is subtracted from all photons

\begin{figure}
	\includegraphics[width=\columnwidth]{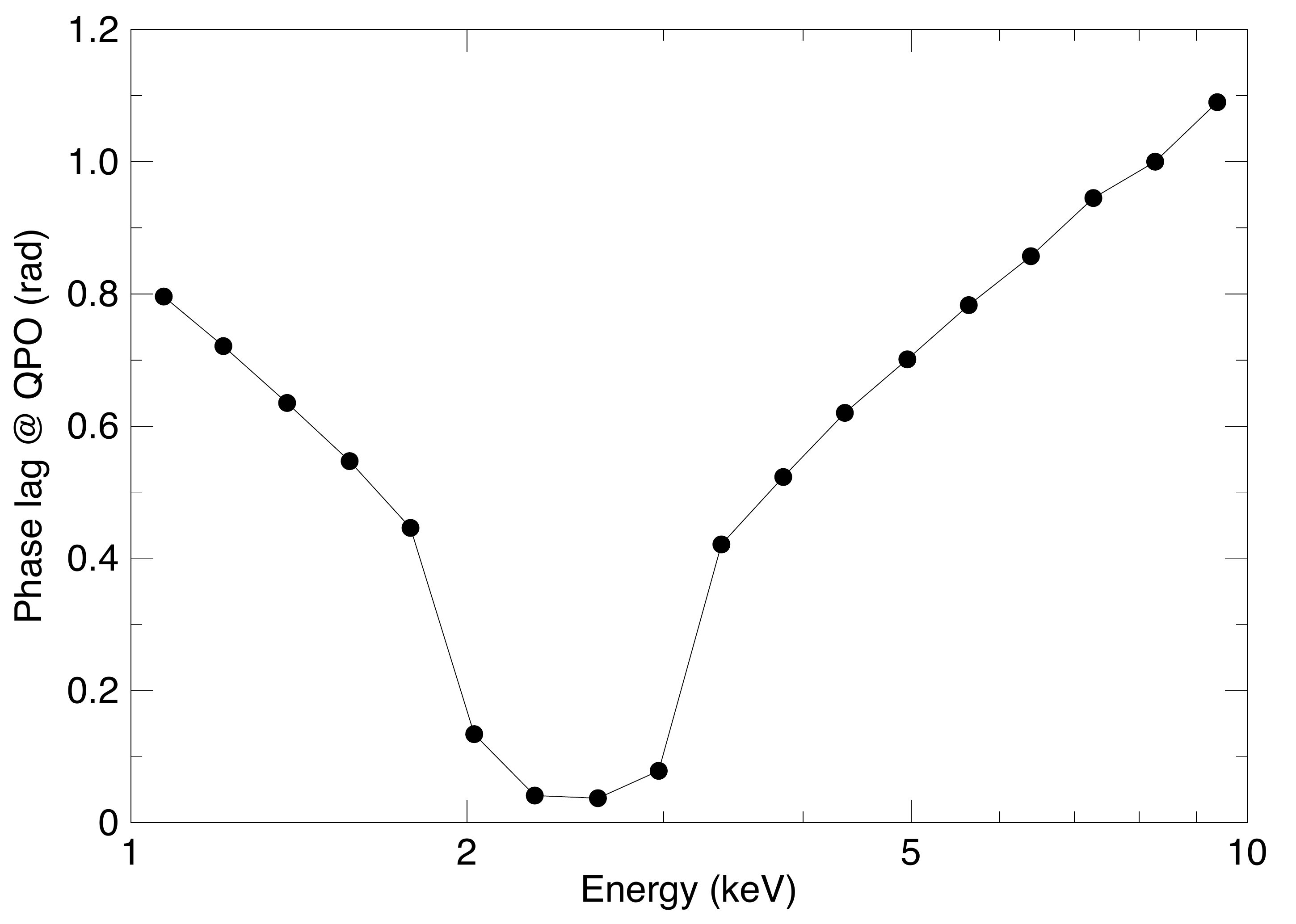}
    \caption{Phase lag at the QPO frequency, as a function of energy, of 2-3 keV simulated input photons that got Comptonised (upscattered and downscattered) in a spherical corona of radius $10^9$ cm, with Thomson depth 0.3 and electron temperature 30 keV.
    }
    \label{fig:nick}
\end{figure}

The results of our calculation are shown in Fig. \ref{fig:nick} and they demonstrate that Comptonization can explain the positive lags at both low and high energies. For the parameters that we have chosen for the hot corona, the computed photon number spectral index $\Gamma$ is approximately 3.5. Clearly, a nonphysically large corona is needed to produce the size of the observed lags, but the simulation shown here is very simplified and it has been used only as a demonstration. A more physical corona is the jet, and Comptonization in the jet \citep{Reig2003,Giannios2005,Markoff2005,Kylafis2008,ReigKylafis2015,KylafisReig2018,Reig2018,ReigKylafis2019} is expected to produce similar results.  This work is currently in progress.

\section*{Acknowledgements}

This research has made use of data and/or software provided by the High Energy Astrophysics Science Archive Research Center (HEASARC), which is a service of the Astrophysics Science Division at NASA/GSFC. TMB acknowledges financial contribution from the agreement ASI-INAF n.2017-14-H.0 and acknowledges fruitful discussion during a Team Meeting at the International Space Science Institute (Bern). LZ acknowledges support from the Royal Society Newton Funds. DA acknowledges support from the Royal Society. 

\section*{Data Availability}

The data underlying this article are available in the HEASARC database, at \url{https://heasarc.gsfc.nasa.gov}.




\bibliographystyle{mnras}
\bibliography{bibliography} 








\bsp	
\label{lastpage}
\end{document}